\newcommand{\tr}{{\rm Tr}}
\def\ra{\rangle}
\def\la{\langle}
\def\be{\begin{equation}}
\def\ee{\end{equation}}
\def\ba{\begin{array}}
\def\ea{\end{array}}

\def\dps{\displaystyle}

\documentclass[aps,amsmath,amssymb,amsfonts,showpacs,onecolumn,prl]{revtex4}

\usepackage{graphicx,color}
\usepackage[toc,page]{appendix}
\usepackage{caption2}

\def\qed{\leavevmode\unskip\penalty9999 \hbox{}\nobreak\hfill
     \quad\hbox{\leavevmode  \hbox to.77778em{%
               \hfil\vrule   \vbox to.675em%
               {\hrule width.6em\vfil\hrule}\vrule\hfil}}
     \par\vskip3pt}

\begin{document}

\title{\large\bf Quantum Discord for $d\otimes2$ Systems}

\author{Zhihao Ma$^{1,2}\footnote{ma9452316@gmail.com}$}
\author{Zhihua Chen$^{3,4}\footnote{chenzhihua77@gmail.com}$}
\author{Felipe Fernandes Fanchini$^{5}\footnote{fanchini@fc.unesp.br}$}
\author{Shao-Ming Fei$^{6,7}$} 
\affiliation{$^1$Department of Mathematics, Shanghai
Jiaotong University, Shanghai 200240, China\\
$^2$Department of Physics and Astronomy, University College London, WC1E 6BT London, United Kingdom\\
$^3$Department of Mathematics, College of Science, Zhejiang University of
Technology, Hangzhou 310023, China\\
$^4$Centre for Quantum Technologies, National University of Singapore, 117543 Singapore\\
$^{5*}$Departamento de F\'isica, Faculdade de Ci\^encias, Universidade
Estadual Paulista\\
$^6$School of Mathematical Sciences, Capital Normal University, Beijing 100048, China\\
$^7$Max-Planck-Institute for Mathematics in the Sciences, 04103 Leipzig, Germany}

\begin{abstract}

We present an analytical solution for classical correlation,
defined in terms of linear entropy, in an arbitrary $d\otimes 2$
system {when the second subsystem is measured}. We show that the optimal measurements used in the
maximization of the classical correlation in terms of linear
entropy, when used to calculate the quantum discord in terms of von
Neumann entropy, result in a tight upper bound for arbitrary
$d\otimes 2$ systems. This bound agrees with all known analytical
results about quantum discord in terms of von Neumann entropy
and, when comparing it with the numerical results for $10^6$ two-qubit
random density matrices, we obtain an average deviation of order
$10^{-4}$. Furthermore, our results give a way to calculate the
quantum discord for arbitrary $n$-qubit GHZ and W states evolving
under the action of the amplitude damping noisy channel.

\end{abstract}


\maketitle

Quantum entanglement plays important roles in many areas of quantum information
processing, such as quantum teleportation and superdense coding
\cite{nielsen,Horodecki09,Guhne09}. Nevertheless, quantum
entanglement is not the only form of quantum correlation that is
useful for quantum information processing. Indeed, some separable
states may also speed up certain quantum tasks,  relative to their classical
counterparts \cite{datta1,datta2,datta3,datta4}, and many quantum tasks, such as
quantum nonlocality \cite{Horodecki09,bennett,niset} and
deterministic quantum computations with one qubit \cite{Knill}, can
be carried out with forms of quantum correlation other than quantum
entanglement. One such quantum correlation, called quantum discord, has
received a great deal of attention recently (see \cite{Modi} and references therein).
Introduced by Ollivier and Zurek \cite{ollivier} as the difference between the quantum mutual information and the maximal conditional
mutual information obtained by local measurements \cite{ollivier,Vedral},
quantum discord plays an important role in some quantum information
processing \cite{ost,LiB}.

Despite much effort by the scientific community, an analytical
solution of quantum discord is still lacking even for two-qubit
systems. Owing to the maximization involved in the calculation, there
are only a few results on the analytical expression of quantum discord
and only for very special states are exact solutions known. However, if
instead of the von Neumann entropy one uses the linear entropy, the
optimal measurements that maximize the conditional mutual
information can be obtained analytically \cite{osborne}. Here, we
show that using these optimal measurements to determine the
quantum discord in terms of the von Neumann entropy results in an
excellent upper bound for the latter. Moreover, we show that this
result gives a way to calculate the quantum discord for arbitrary
$n$-qubit GHZ and W states, with each qubit subjected to the
amplitude damping channel individually.

\section{Results}
\subsection{Classical correlation under linear entropy}

The usual quantum discord, in terms of von Neumann entropy, is defined as follows:
let $\rho^{AB}$ denote the density operator of a bipartite
system composed of partitions $A$ and $B$.  Let
$\rho^{A}=Tr_{B}(\rho^{AB})$ and $\rho^{B}=Tr_{A}(\rho^{AB})$ be the
reduced density operators of $A$ and $B$, respectively. The quantum
mutual information, which is the information-theoretic
measure of the total correlation, is defined as
$\mathcal{I}(\rho^{AB})=S(\rho^A)+S(\rho^B)-S(\rho^{AB})$, where
$S(\rho)=-\rm{Tr}(\rho\log_2\rho)$ is the von Neumann entropy.
Usually, the total correlation $\mathcal{I}$ is split into the quantum part
$\mathcal{Q}$ and the classical part $\mathcal{C}$, such that
$\mathcal{I}=\mathcal{Q}+\mathcal{C}$. The classical correlation of
a bipartite state $\rho^{AB}$ is defined as
\begin{equation}
\mathcal{C}(\rho^{AB})=I^{\leftarrow}(\rho^{AB})=\max\limits_{P_i}[S(\rho^{A})-\sum_{i}p_{i}S(\rho^{A|i})],\label{def:classical correlation}
\end{equation}
{where the maximum is taken over all positive operator-valued
measurements (POVM) $\{P_i\}$ performed on subsystem $B$, satisfying $\sum_i  P_i^{\dagger}P_i =1$, with
probability of $i$ as an outcome, $p_i\equiv \tr[({\rm I}^{A}\otimes P_i)\rho^{AB}({\rm I}^{A}\otimes P_i^\dagger)]$ where $\rho^{A|i}\equiv \tr_B[({\rm I}^{A}\otimes {P_i})\rho^{AB}({\rm I}^{A}\otimes P_i^\dagger)]/p_i$ is
the conditional state of system $A$ associated with outcome $i$, where
${\rm I}^{A}$ is the identity operator on subsystem $A$. }


In this work, all POVM or projective measurements (PM) are taken on subsystem B.
Finally, the quantum discord is defined as the difference
between the total correlation and the classical correlation \cite{ollivier,Vedral}:
\be\label{def:quantum correlation}
\ba{rcl}
\mathcal{Q}(\rho^{AB})&=&\mathcal{I}(\rho^{AB})-I^{\leftarrow}(\rho^{AB})\\[2mm]
&=&\displaystyle\min_{P_i}\sum_{i}p_i S(\rho^{A|i})-S(\rho^{A|B}),
\ea
\ee
where $S(\rho^{A|B})=S(\rho^{AB})-S(\rho^{B})$ is the conditional entropy.

To calculate our tight upper bound to quantum discord, instead of the von Neumann entropy one uses the linear entropy. The linear entropy of a state $\rho$ is given by:
\begin{equation}
S_{2}(\rho)=2[1-\rm{Tr}(\rho^{2})].\label{linear-entropy}
\end{equation}
In terms of the linear entropy (\ref{linear-entropy}), one can
correspondingly define the conditional linear entropy, $S_{2}(A|B)=S_{2}(\rho^{AB})-S_{2}(\rho^B)$,
and the classical correlation \cite{osborne} is written as:
\begin{equation}
I_{2}^{\leftarrow}(\rho^{AB})=\max\limits_{P_i}[S_{2}(\rho^{A})-\sum\limits_{i}p_{i}S_{2}(\rho^{A|i})],\label{linear-entropy-cc}
\end{equation}
where the measurements run over all POVMs $P_i$.

Although the classical correlation and, consequently, the quantum discord (\ref{def:quantum correlation})
is extremely difficult to compute in terms of von Neumann entropy, the classical correlation (\ref{linear-entropy-cc}) expressed in terms of linear entropy
can be calculated analytically. In what follows we present the analytical formula for an arbitrary $d\otimes 2$ quantum systems.

A qudit state can be written as $\rho=({\rm
I}_{d}+{{\mathbf{r}\cdot\mathbf{\gamma}}})/d$, where ${\rm
I}_d$ denotes the $d\times d$ identity matrix, $\mathbf{r}$ is a
$(d^{2}-1)$-dimensional real vector,
{{$\mathbf{\gamma}=(\gamma_{1},\gamma_{2},...,\gamma_{d^{2}-1})^{T}$}} is
the vector of generators of $SU(d)$ and $T$ stands for transpose.
Consider a bipartite system, composed of a $d$-dimensional subsystem
labeled  $A$ and a $2$-dimensional subsystem labeled $B$. The
bipartite state $\rho^{AB}$ can be written as:
\begin{equation}
\rho^{AB}=\Lambda\otimes {{ \mathbb{1}}}(|V_{B'B}\rangle\langle V_{B'B}|),\label{J-Choi-pure}
\end{equation}
where $|V_{B'B}\rangle$ is the symmetric two-qubit purification of
the reduced density operator $\rho^B$ on an auxiliary qubit system
$B'$ and $ \mathbb{1}$ is the identity map on system $B$. Here, symmetric two-qubit purification means that the two
reduced density matrices are equal, i.e. $V_{B'}=V_{B}=\rho^B$, and $\Lambda$ is a a completely positive trace-preserving map which maps a qubit state $B'$
to the qudit state $A$.
Let {{$\mathbf{\sigma}=(\sigma_{1},\sigma_{2},\sigma_{3})^{T}$}}
denote the vector of Pauli operators, $\mathbf{r}$ being a three-dimensional vector,
$|\mathbf{r}|\leq 1$. As a qubit state can generally be written as
$\rho=(\rm{I}_d+\mathbf{r}\cdot{{\mathbf{\sigma}}})/d$, the map $\Lambda(\rho)$ is of the form
\begin{equation}
\Lambda(\rho)=[\rm{I}_d+(\mathbf{L}\mathbf{r}+\mathbf{s})\cdot \mathbf{\gamma}]/{d}, \label{channel00}
\end{equation}
where $\mathbf{L}$ is a $(d^{2}-1)\times3$ real matrix  {{and $\mathbf{s}$ is a three-dimensional vector}. $\mathbf{L}$ and $\mathbf{s}$
can be obtained from Eq. (\ref{J-Choi-pure}) and Eq. (\ref{channel00}). Let $\rho_B=\sum\limits_0^1\lambda_i|\phi_i\rangle\langle\phi_i|$ be the spectral
decomposition of $\rho_B$. Then $|V_{B'B}\rangle=\sum\limits_{i=0}^1\sqrt{\lambda_i}|\phi_i\rangle|\phi_i\rangle$ and $\Lambda(|i\rangle\langle j|)$, $i,j=0,1$,
can be calculated by Eq. (\ref{J-Choi-pure}). Therefore one gets $\Lambda(\sigma_i)$, $i=1,2,3$, and the matrix
$L_{i,j}=\tr(\Lambda(\sigma_j).\sigma_i)$.} By the method used to calculate
the classical correlation $I_{2}^{\leftarrow}(\rho^{AB})$ of two-qubit states \cite{osborne}, we have:
\begin{equation}\label{T1}
I_{2}^{\leftarrow}(\rho^{AB})=\lambda_{max}(\mathbf{L}^{T}\mathbf{L}) S_{2}(\rho^B),
\end{equation}
where $\lambda_{max}(\mathbf{L}^{T}\mathbf{L})$ stands for the largest eigenvalue of
the matrix $\mathbf{L}^{T}\mathbf{L}$. Eq. (\ref{T1}) gives the analytical formula for the classical
correlation in terms of linear entropy for a general $d\otimes 2$
quantum state. Indeed, one only needs to find the eigenvalues of the matrix
$\mathbf{L}^{T}\mathbf{L}$.

Since, for a given state $\rho^{AB}$, the reduced state $\rho^B$,
$|V_{B'B}\rangle$ and the map $\Lambda$ are fixed, the classical correlation
can readily be computed in terms of linear entropy $I_2^\leftarrow(\rho^{AB})$. What concern us here are the optimal measurements
that give rise to $I_2^\leftarrow(\rho^{AB})$. In fact,
there is a one-to-one correspondence between all possible
POVM measurements and all convex decompositions of $\rho^{B}$
\cite{Hughston93}; namely, if
$\rho^{B}=\sum\limits_{j=0}^{1}p_{j}|\psi_{j}\ra\la \psi_{j}|$ is
the pure state decomposition of $\rho^{B}$, then the following
are the corresponding POVMs:
\begin{eqnarray}
M_0=(\rho^B)^{-\frac{1}{2}}p_0 |\psi_{0}\ra\la \psi_{0}| (\rho^B)^{-\frac{1}{2}},\\[2mm]
M_1=(\rho^B)^{-\frac{1}{2}}p_1 |\psi_{1}\ra\la \psi_{1}| (\rho^B)^{-\frac{1}{2}},
\end{eqnarray}
where $\rho^{B}$ is full-ranked.
Otherwise, we can find the inverse of $\rho^B$ in its range
projection and, from the optimal pure state decompositions of $\rho^{B}$, we
can get the corresponding optimal POVMs. In \cite{osborne}, the authors have shown
how to find the optimal decomposition of $\rho^{B}$. First write
$\rho^{B}$ in its Bloch form:
$\rho^{B}=(I+\mathbf{r_{B}}\cdot\vec{\sigma})/2$.
Let {{$\mathbf{r_{B}}+X_{j}$ be the Bloch vector for the
pure state decomposition $|\psi_{j}\ra$ of $\rho^{B}$,
where $X_{j}=(X^{x}_{j},X^{y}_{j},X^{z}_{j})$ and $\sum\limits_j p_j X_j=0$}},
$\sum\limits_{j}p_{j}(\mathbf{r_{B}}+ X_{j})=\mathbf{r_{B}}$. Hence,
$||\mathbf{r_{B}}+X_{j}||=1$.{{Then $(X_j^x)^2=1-||\mathbf{r_B}||^2-2 \mathbf{r_B}^T X_j-(X_j^y)^2-(X_j^z)^2$. Without loss of generality, assume that $L^TL$ is diagonal with diagonal elements $\lambda_x\geq \lambda_y\geq \lambda_z,$ Eq. (\ref{T1}) becomes $\lambda_x(1-||\mathbf{r_B}||^2)+\max\limits_{p_j,X_j} p_j[(\lambda_y-\lambda_x)(X_j^y)^2+(\lambda_z-\lambda_x)(X_j^z)^2]$,
which gets the maximum value when $X_j^y=X_j^z=0$. There are exactly two solutions of the
equation $||\mathbf{r_{B}}+X_{j}||=1$. Hence the optimal
decomposition of $\rho^{B}$ reads:
$\rho^{B}=\sum\limits_{j=0}^{1}p_{j}|\psi_{j}\ra\la \psi_{j}|$.}}
From the two pure states in
the optimal decomposition, we obtain the two
optimal POVM measurement operators $M_{0}$ and $M_{1}$.

{It is well known that to maximize the classical correlation it is
necessary to use the most general POVM quantum measurement.  As it
is much more complicated to find the maximum in (\ref{def:classical
correlation}) over all POVMs than over von Neumann measurements,
almost all known analytical results are based on the latter. Indeed,
only very few results are based on POVM \cite{Shi12,Chen11}. Here,
we show that for the case of a bipartite qudit-qubit state, the
classical correlation based on linear entropy is maximized
over projective measurements (see proof in the appendix).
This leads to our first theorem:}

{\bf Theorem 1.} The classical correlation of a qudit-qubit state
$\rho_{AB}$ defined  by running over all (arbitrary) POVM measurements
is the same as the classical correlation defined by running
over all projective measurements, i.e.,
$I^{\leftarrow}_{2\,\rm{POVM}}(\rho^{AB}) = I^{\leftarrow}_{2\,{{\rm{PM}}}}(\rho^{AB})$.

\subsection{Quantum discord under von Neumann entropy}
Theorem 1 implies that the optimal POVM in the classical correlation defined
by Eq. (\ref{linear-entropy-cc}) is in fact a projective
measurement. This is very different from the case of classical
correlation $I^{\leftarrow}(\rho^{AB})$ defined by von Neumann
entropy, in which the classical correlation based on POVM could be
larger than the one based on projective measurement
\cite{Shi12,Chen11}. This shows that,
although von Neumann entropy and linear entropy 
have many properties in common,
they behave quite differently in optimizing classical information.
However, by using the optimal projective measurement for the
classical correlation $I_{2}^{\leftarrow}(\rho^{AB})$ based on
linear entropy, we can get a tight lower bound for the classical
correlation based on von Neumann entropy, and hence a tight upper
bound for the quantum discord based on von Neumann entropy. This leads us to our second theorem:

{\bf Theorem 2.} The quantum discord based on von Neumann entropy has an upper bound:
\begin{equation}\label{lb}
\mathcal{Q}(\rho^{AB})\leq \mathcal{I}(\rho^{AB})- [S(\rho^{A})-\sum\limits_{i}p_{i}S(\rho^{A|i})],
\end{equation}
where $p_i\equiv \tr[(I^{A}\otimes P_i)\rho^{AB}(I^{A}\otimes P_i^\dagger)]$
is the probability of the measurement outcome $i$,
$\rho^{A|i}\equiv \tr_B[(I^{A}\otimes {P_i})\rho^{AB}(I^{A}\otimes P_i^\dagger)]/p_i$ is the conditional state
of system A when the measurement outcome is $i$,
and $P_0$ and $P_1$ are the optimal projective measurement
operators for $I_{2}^{\leftarrow}(\rho^{AB})$ of a given $d\otimes 2$ state $\rho_{AB}$.

{{In fact, there is a connection between discord and entanglement of formation (EOF):
the classical correlation $I^{\leftarrow}(\rho^{AB})$ can be obtained from EOF by the Koashi Winter Relation \cite{Koashi},
\begin{equation}\label{lc}
I^{\leftarrow}(\rho^{AB})+E(\rho_{AC})=S(\rho^A),
\end{equation}
where $I^{\leftarrow}(\rho^{AB})$ is the original classical correlation of $n\otimes 2$ state $\rho^{AB}$, $E(\rho_{AC})$ is the EOF of state $\rho^{AC}$, and $\rho^{AC}$
is the purification of $\rho^{AB}$. It is important to note that, from theorem 2, we can get an upper bound of EOF for arbitrary rank two $n\otimes m$ state $\rho^{AC}$.}}

Although the upper bound (\ref{lb}) of $\mathcal{Q}(\rho^{AB})$ is
given by the optimal measurement of $I_2^\leftarrow(\rho^{AB})$, we
show, by means of examples, that it is a surprisingly good estimate of
$\mathcal{Q}(\rho^{AB})$.

{\it Example 1.} In \cite{Luo} Luo presented the analytic formula
for the quantum discord $\mathcal{Q}(\rho^{AB})$ of the two-qubit
Bell-diagonal state:
$\rho=(I\otimes I+\sum_{i=1}^{3}c_{i}\sigma_{i}\otimes \sigma_{i})/4$.
Let $c=max\{|c_1|,|c_2|,|c_3|\}$. For this Bell-diagonal state, {{$L^TL=\mathbf{Diag}\{c_1^2,c_2^2,c_3^2\}$ and $\mathbf{r_B}=0$.
The two solutions of $X_j$ are $(1,0,0)^T$ and $(-1,0,0)^T$ when $c_1^2\geq \max\{c_2^2,c_3^2\}$: $|\psi_0\rangle=\frac{1}{2}(I+\sigma_1)$ and $|\psi_1\rangle=\frac{1}{2}(I-\sigma_1)$;
$(0,1,0)^T$ and $(0,-1,0)^T$ when $c_2^2\geq \max\{c_1^2,c_3^2\}$: $|\psi_0\rangle=\frac{1}{2}(I+\sigma_2)$ and $|\psi_1\rangle=\frac{1}{2}(I-\sigma_2)$;
$(0,0,1)^T$ and $(0,0,-1)^T$ when $c_3^2\geq \max\{c_1^2,c_2^2\}$: $|\psi_0\rangle=\frac{1}{2}(I+\sigma_3)$ and $|\psi_1\rangle=\frac{1}{2}(I-\sigma_3)$. }}
It can be verified immediately that
the optimal measurements for $I_{2}^{\leftarrow}(\rho^{AB})$ are given by
$M_1=(I+\sigma_k)/2$ and $M_2=(I-\sigma_k)/2$, for $c=c_k$ with $k=\{1,2,3\}$.
It can easily be checked that our upper bound (\ref{lb}) is exactly the same as the analytical
results in \cite{Luo}.

{\it Example 2.} In \cite{Li,x-states1,x-state2} the X-type two-qubit states are investigated:
$\rho_{X}=(I\otimes I +x_3(\sigma_3\otimes I)+y_3(I\otimes
\sigma_3)+\sum_{i=1}^3 t_i(\sigma_i\otimes\sigma_i))/4$,
where $x_3$, $y_3$, $t_1$, $t_2$ and $t_3$ are defined such that
$\rho$ is a quantum state. It can easily be seen that our upper bound
(\ref{lb}) agrees perfectly with the analytical results obtained in \cite{Li} (see Fig. 1).

\begin{figure}[htb]
  \centering
  \includegraphics[width=5.5cm]{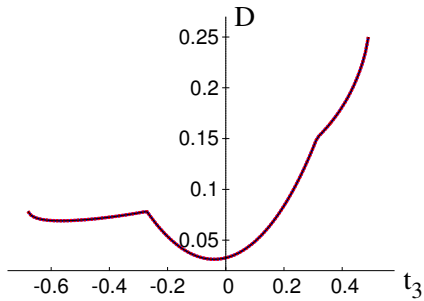}
	\caption{Quantum discord $D=\mathcal{Q}(\rho_{X})$ for $x_3=0.1$, $y_3=0.2$, $t_1=0.2$, $t_2=0.3$. Here the results in \cite{Li}, our numerical results and our upper bound in Eq. (\ref{lb}) agree with high precision.}
\end{figure}
	
Now, let us consider the following general two-qubit states,
$\rho_{2\times 2}=(I\otimes I+\sum_{i=1}^3 [x_i(\sigma_i\otimes
I)+y_i(I\otimes \sigma_i)+t_i(\sigma_i\otimes\sigma_i)])/4$,
and compare our analytical upper bound with numerical results. {{Fig.2}} gives the quantum discord $\mathcal{Q}(\rho_{2\times 2})$, for $x_1=0.05$,
$x_2=0.1$, $x_3=0.1$, $y_1=0.15$, $y_2=0.25$, $y_3=0.2$, $t_1=0.2$
and $t_2=0.2$ plotted against $t_3$, such that $\rho_{2\times 2}$ is a quantum
state. {{Fig.3}} shows the quantum discord $\mathcal{Q}(\rho_{2\times 2})$ for
$x_2=0.1$, $x_3=0.1$, $y_1=0.15$, $y_2=0.25$, $y_3=0.2$, $t_1=0.2$,
$t_2=0.2$ and $t_3=-0.5$, plotted against $x_1$, such that $\rho_{2\times 2}$ is
a quantum state. It can be seen that our upper bound coincides very well with the numerical results.

\begin{figure}[htb]
  \centering
  \includegraphics[width=11 cm]{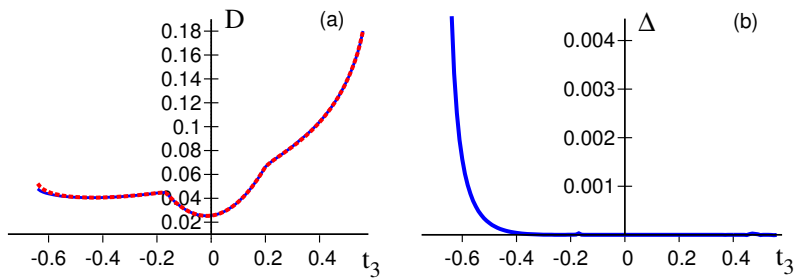}
	\caption{Figure (a) shows quantum discord $D=\mathcal{Q}(\rho_{2\times 2})$. Solid blue line shows numerical results and the red dotted line our upper bound. Figure (b) shows the difference between the numerical results and our upper bound.}
\end{figure}
	
\begin{figure}[htb]
  \centering
  \includegraphics[width=11 cm]{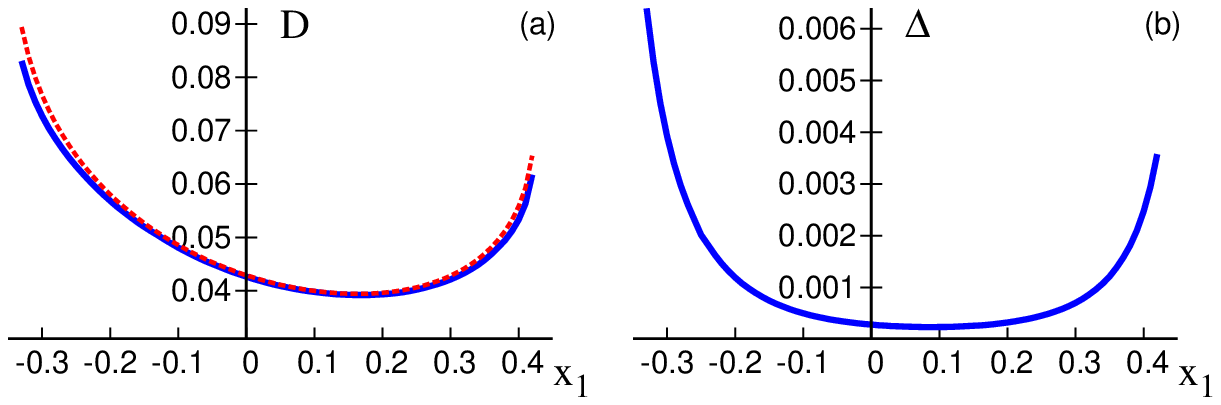}
	\caption{Figure (a) shows quantum discord $D=\mathcal{Q}(\rho_{2\times 2})$. Solid blue line shows numerical results and the red dotted line our upper bound. Figure (b) shows the difference between the numerical results and our upper bound.}
\end{figure}

\begin{figure}
\includegraphics[width=8 cm]{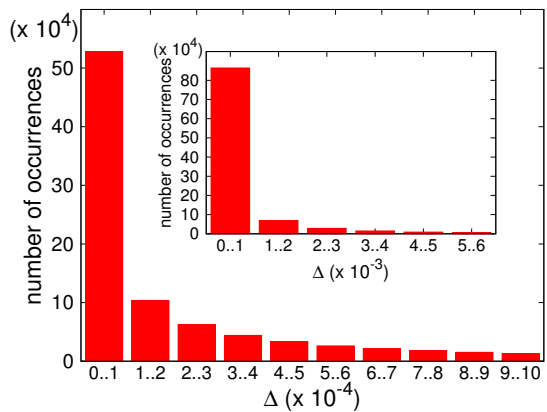}
\caption{$\Delta$ as a function of number of occurrences for a set of $10^6$ random {{$2\otimes 2$}} density matrices.}
\end{figure}

We have seen that the upper bound of quantum discord based on von Neumann entropy,
obtained from the optimal measurements for the classical correlation based on linear entropy,
is often exact. To test the precision of our upper bound generally, we
calculated the difference between our analytical result and the
numerical solution of quantum discord, for a set of $10^6$ random
density matrices {{of $2\otimes 2$}}. In Fig.{{4}}, we plot the deviation
$\Delta=\mathcal{Q}_{\rm{Analytical}}-\mathcal{Q}_{\rm{Numerical}}$
against the number of occurrences. It can be seen that more than
half of the randomly generated density matrices results in a precision
greater than $10^{-4}$, which demonstrates that our analytical
result is a tight upper bound. Furthermore, in Fig.{{4}}, we show
that more than $80\%$ of the density matrices randomly generated
lead to a precision greater than $10^{-3}$. Indeed, the percentage of
density matrices with a deviation $\Delta$ greater than
$6\times10^{-3}$ is less than $0.1\%$. {{Here, in the horizontal coordinate of Fig. 4,  $0..1$ represents the interval from 0 to 1, and the same for $1..2$, etc.}}.

\subsection{Evolution of Quantum Discord under AD Channel}
Now we consider the evolution of quantum discord for arbitrary $n$-qubit
GHZ and W states under an amplitude damping (AD) channel characterized by
the Kraus operators $E_1=\sqrt{p}\left(\ba{cc}0&1\\0&0\ea\right)$
and $E_2=\left(\ba{cc}1&0\\0&\sqrt{1-p}\ea\right)$.
We show that the related quantum discord based on von Neumann entropy
can be analytically obtained from the upper bound given by Eq. (\ref{lb}).

First let us consider $n$-qubit GHZ states, with the first $(n-1)$ qubits subjected to AD
channels individually. From Theorem 2, we get the optimal measurement operators
$(I_2+\sigma_z)/2$ and $(I_2-\sigma_z)/2$ for classical correlation in terms of linear entropy, and the upper bound of quantum discord
in terms of von Neumann entropy is then exact.
Let $M_1=U\,P_0\,U^{+}$ and $M_2=U\,P_1\,U^{+}$ be the two measurement operators, where
$P_0$ and $P_1$ are the projective operators, $U=t I_2+y_1\sigma_x+y_2\sigma_y+y_3\sigma_z$
with $t^2+y_3^2=\cos^2\theta$ and $y_1^2+y_2^2=\sin^2\theta$.
Fig.{{5}} shows that when $\theta=0$ or $\theta=\pi,$ $\sum_{i}p_{i}S(\rho^{A|i})$
has the minimal value, which coincides with the optimal measurement operators
$(I_2+\sigma_z)/2$ and $(I_2-\sigma_z)/2$ for classical correlation based on linear entropy.

\begin{figure}
\includegraphics[width=0.32\textwidth]{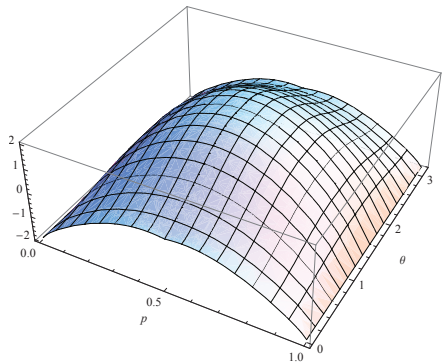}
\caption{$\sum\limits_{i}p_{i}S(\rho^{A|i})$ as a function of $\theta$ and p.}
\end{figure}

For $n$-qubit W states with the first $(n-1)$ qubits subjected to individual AD
channels, from Theorem 2 we have the optimal measurement
operators $(I_2+\sigma_x)/2$ and $(I_2-\sigma_x)/2$ or
$(I_2+\sigma_y)/2$ and $(I_2-\sigma_y)/2$. The upper bound of
quantum discord obtained in terms of these measurement operators
coincide with its lower bound in \cite{tightbound}.
It follows that again  we have the exact value of quantum discord (\ref{def:quantum
correlation}).

Alternatively, if the last qubit of an $n$-qubit W state is subjected to an AD channel,
we have the optimal measurement operators $(I_2+\sigma_x)/2$ and
$(I_2-\sigma_x)/2$ or $(I_2+\sigma_y)/2$ and
$(I_2-\sigma_y)/2$, which also give rise to the exact value of discord (\ref{def:quantum correlation}).

\section{Conclusions}
We have studied the quantum discord of
qudit-qubit states. The analytical formula for classical correlation
based on linear entropy has been explicitly derived, from which an
analytical tight upper bound of quantum discord based on von Neumann
entropy is obtained for arbitrary qudit-qubit states. The upper
bound is found to be surprisingly good in the sense that it agrees
very well with all known analytical results about quantum
discord in terms of von Neumann entropy. Furthermore, for a set of
$10^6$ random density matrices, the maximum deviation found from the
numerical solution was approximately $0.05$ and the number of
density matrices whose deviation was greater than $6\times10^{-3}$ was
less than $0.1\%$ of the whole set. Our analytical results could be
used to investigate the roles played by quantum discord in quantum
information processing. For classical correlation in terms of linear
entropy, it has also been shown that the result for a qudit-qubit
state, defined by running over all two-operator POVM measurements,
is equivalent to that defined by running over all projective
measurements. Furthermore, our results can be applied to investigate
the evolution of quantum discord for arbitrary $n$-qubit GHZ and W
states. Indeed, employing an important paradigmatic noisy channel,
we present the quantum discord dynamics for the GHZ and W states when
each qubit is subjected to independent amplitude damping channels.

\section{Acknowledgements}
The work is supported by NSFC under
numbers 11371247, 10901103 and 11201427. FFF is supported by S\~ao
Paulo Research Foundation (FAPESP), under grant number 2012/50464-0,
and by the National Institute for Science and Technology of Quantum
Information (INCT-IQ), under process number 2008/57856-6. FFF is also
supported by the National Counsel of Technological and Scientific
Development (CNPq) under grant number 474592/2013-8.

\section{Author contributions}
Z.M. and S.F. prove the main theorems, Z.C. and F.F.F. developed the numerical codes, and Z.M., Z.C., F.F.F. and S.F. wrote the manuscript.

\smallskip
\textbf{Competing financial interests:} The authors declare no competing financial interests.

\appendix
\section{Appendix}
\noindent{\sf [Proof of Theorem 1]}~~ Theorem 1 can be proved by using an approach similar
to that used in \cite{superdiscord}. It was proved by \cite{osborne}, for the classical correlation $I^{\leftarrow}_{2\,\rm{POVM}}(\rho^{AB})$,  of a qudit-qubit state
$\rho_{AB}$ defined  by running over all POVM measurements(here all POVM means we run over arbitrary POVM, that is, any measurement operators POVM), its optimal POVM measurement must be two operators POVM, that is, we can in fact restrict to two operators POVM. Then let $M(x)$ and $M(-x)$ be two optimal POVM
operators, such that $M(x)+ M(-x)=I$. Let $\Pi_{0}$ and $\Pi_{1}$ be the
projective measurement operators, $\Pi_{0}+ \Pi_{1}=I$.
We can write the two POVM operators as
$M(x)=\sum_{i=0}^{1}a_i(x)\Pi_{i}$ and $ M(-x) =
\sum_{i=0}^{1}a_i(-x)\Pi_{i}$, where $a_0(x)$ and $a_1(x)$ are the
eigenvalues of $M(x)$ and $a_0(-x)$ and $a_1(-x)$ are eigenvalues of
$M(-x)$.

Given any qudit-qubit state $\rho_{AB}$, the POVM  $\{M(x),M(-x)\}$ performed on subsystem $B$ will yield the
post-measurement state $\rho_{A|M^{B}(x)}$. We have
$p(x)\rho_{A|M^{B}(x)}=\mbox{Tr}_B[\rho_{AB}(I \otimes M(x))]
=\sum_{i=0}^{1} a_i(x)\,p_{i}\,\rho_{A|\Pi_i^B}$.
From the concavity property of linear entropy, we have the lower
bound of the  conditional linear entropy,
$$\ba{l}
\dps\sum_{y=x,-x}p(y)S_{2}(\rho_{A|M^{B}(y)})\\
\dps=\sum_{y=x,-x}p(y) S_{2}(\sum_{i=0}^{1} \frac{a_i(y)p_{i}}{p(y)}\rho_{A|\Pi_i^B})\\
\dps\geq\sum_{y=x,-x}p(y) \sum_{i=0}^{1} \frac{a_i(y)p_{i}}{p(y)}S_{2}(\rho_{A|\Pi_i^B})=\sum_{i=0}^{1}p_{i}S_{2}(\rho_{A|\Pi_i^B}).
\ea
$$
Thus, the conditional linear entropy derived from POVMs is greater than or
equal to the conditional linear entropy derived from the projective
measurements, on the all possible measurements basis.

Let $\{ {\tilde \Pi}_i^B \}$ be the measurement basis that maximizes
the classical correlation $I^{\leftarrow}_{2\,\rm{POVM}}(\rho^{AB})$ for two POVM operators $\{M(x),M(-x)\}$.
Then, 
we have
$$
\ba{l}
\dps I^{\leftarrow}_{2\,\rm{POVM}}(\rho^{AB})=S_{2}(\rho^{A})-\sum_{y=x,-x}p(y)S_{2}(\rho_{A|M^{B}(y)})\\[4mm]
~~\dps \leq S_{2}(\rho^{A})-\sum_{i}p_{i}S_{2}(\rho_{A|\Pi_i^B})
\dps \leq I^{\leftarrow}_{2\,\rm{PM}}(\rho^{AB}),
\ea
$$
since $\{ {\tilde \Pi}_i^B \}$ could be a non-optimal projective
measurement of the classical correlation
$I^{\leftarrow}_{2\,\rm{PM}}(\rho^{AB})$. Hence,
$I^{\leftarrow}_{2\,\rm{POVM}}(\rho^{AB}) \leq
I^{\leftarrow}_{2\,\rm{PM}}(\rho^{AB})$; i.e., the classical
correlation under two POVM measurements is always smaller than or
equal to the classical correlation under projective measurements.

On the other hand,  a projective measurement is a  POVM. Hence, by
definition, the classical correlation under two POVM measurements is
always greater than or equal to the classical correlation under
projective measurement. Therefore, we have proved that the classical
correlation under two POVM measurements is equal to the classical
correlation under projective measurement,
$I^{\leftarrow}_{2\,\rm{POVM}}(\rho^{AB}) =
I^{\leftarrow}_{2\,\rm{PM}}(\rho^{AB})$. $\hfill\Box$


\begin{thebibliography}{10}
\bibitem{nielsen} Nielsen M. A. \& Chuang I. L. \emph{Quantum Computation and Quantum Information}
          (Cambridge University Press, Cambridge, England, 2000).
          
\bibitem{Horodecki09} Horodecki R.,  Horodecki P., Horodecki M., \& Horodecki K. Quantum entanglement. \emph{Rev. Mod. Phys.} \textbf{81}, 865 (2009).

\bibitem{Guhne09} G\"uhne O. \&  T\'oth G. Entanglement detection. \emph{Phys. Rep.} \textbf{474}, 1 (2009).

\bibitem{datta1} Datta A., Flammia A. T., \& Caves C. M. Entanglement and the power of one qubit. \emph{Phys. Rev. A} \textbf{72}, 042316 (2005);

\bibitem{datta2} Datta A. \&  Vidal G. Role of entanglement and correlations in mixed-state quantum computation. \emph{Phys. Rev. A} \textbf{75}, 042310 (2007);

\bibitem{datta3} Datta A., Shaji A. \&  Caves C. M. Quantum discord and the power of one qubit. \emph{Phys. Rev. Lett.} \textbf{100}, 050502 (2008);

\bibitem{datta4} Lanyon B. P., Barbieri M., Almeida M. P. \& White A. G. Experimental quantum computing without entanglement. \emph{Phys. Rev. Lett.}  \textbf{101}, 200501 (2008).

\bibitem{bennett} Bennett C. H. et al. Quantum nonlocality without entanglement. \emph{Phys. Rev. A} \textbf{59}, 1070 (1999).

\bibitem{niset} Niset J. \&  Cerf N. J. Multipartite nonlocality without entanglement in many dimensions.  \emph{Phys. Rev. A.} \textbf{74}, 052103 (2006).

\bibitem{Knill}  Knill E. \&  Laflamme R. On the power of one bit of quantum information. \emph{Phys. Rev. Lett.} {\bf 81}, 5672 (1998).

\bibitem{Modi} Modi K.,  Brodutch A., Cable H.,  Paterek T. \&  Vedral V. The classical-quantum boundary for correlations: discord and related measures. \emph{Rev. Mod. Phys.} \textbf{84}, 1655 (2012).

\bibitem{ollivier} Ollivier H. \& Zurek W. H. Quantum discord: a measure of the quantumness of correlations. \emph{Phys. Rev. Lett.} \textbf{88}, 017901 (2001).

\bibitem{Vedral} Henderson L. \&  Vedral V. Classical, quantum and total correlations. \emph{J. Phys. A} {\bf 34}, 6899 (2001).

\bibitem{ost} Roa L.,  Retamal J. C. \&  Alid-Vaccarezza M., Dissonance is required for assisted optimal state discrimination. \emph{Phys. Rev. Lett.} \textbf{107}, 080401 (2011);

\bibitem{LiB} Li B.,  Fei S. M., Wang Z. X. \&  Fan H. Assisted state discrimination without entanglement.  \emph{Phys. Rev. A} \textbf{85}, 022328 (2012).

\bibitem{osborne} Osborne T. J. \&  Verstraete F. General monogamy inequality for bipartite qubit entanglement. \emph{Phys. Rev. Lett.} {\bf 96}, 220503 (2006).

\bibitem{LOO1} Yu S. \& Liu N. Entanglement detection by local orthogonal observables. \emph{Phys. Rev. Lett.} 95, 150504 (2005); 

\bibitem{LOO2} Hassan A. S. M. \&  Joag P. S. Separability criterion for multipartite quantum states based on the Bloch representation of density matrices. \emph{Quant. Inf. and Comp.}  \textbf{8},  0773 (2008).



\bibitem{Hughston93} Hughston L. P.,  Jozsa R. \& Wootters W. K. A complete classification of quantum ensembles having a given density matrix. \emph{Phys. Lett. A} 183, 14 (1993).

\bibitem{Shi12}Shi M., Sun C., Jiang F.,  Yan X. \& Du J. Optimal measurement for quantum discord of two-qubit states. \emph{Phys. Rev. A} \textbf{85}, 064104 (2012).

\bibitem{Chen11} Chen Q.,  Zhang C.,  Yu S.,  Yi X. X. \&  Oh C. H. Quantum discord of two-qubit X states. \emph{Phys. Rev. A} \textbf{84}, 042313 (2011).

\bibitem{Luo}Luo S. Quantum discord for two-qubit systems. \emph{Phys. Rev. A} \textbf{77}, 042303 (2008).

\bibitem{Koashi} Koashi M. \& Winter A. Monogamy of entanglement and other correlations. \emph{Phys. Rev. A} 69, 022309(2004).

\bibitem{Li}Li B., Wang Z. \&  Fei S. Quantum discord and geometry for a class of two-qubit states. \emph{Phys. Rev. A} \textbf{83}, 022321 (2011).

\bibitem{x-states1} Ali M., Rau A. R. P. \&  Alber G. Quantum discord for two-qubit X states. \emph{Phys. Rev. A} \textbf{81}, 042105 (2010); 

\bibitem{x-states2} Ali M., Rau A. R. P. \&  Alber G. Erratum: quantum discord for two-qubit X states. \emph{Phys. Rev. A}  \textbf{82}, 069902 (2010).


\bibitem{Sai} Vinjanampathy S. \&  Rau A. R. P. Quantum discord for qubit-qudit systems. \emph{J. Phys. A: Math. Theor.} \textbf{45}, 095303 (2012).

\bibitem{tightbound}Yu S. X., Zhang C. J., Chen Q. \& Oh C.H. Tight bounds for the quantum discord. \emph{arXiv:} 1102.1301.


\end{thebibliography}

\begin{thebibliography}{10}
\bibitem{superdiscord} Singh U. \&  Pati A. K. Super quantum discord with weak measurements. \emph{Ann. of Phys.} \textbf{343}, 141 (2014).
\bibitem{osborne} Osborne T. J. \&  Verstraete F. General monogamy inequality for bipartite qubit entanglement. \emph{Phys. Rev. Lett.} {\bf 96}, 220503 (2006).
\end{thebibliography}
\end{document}